\documentclass[journal=jcisd8, manuscript=article, layout=onecolumn]{achemso}

\usepackage{graphicx}
\usepackage{booktabs}
\usepackage{subcaption}   
\usepackage{amsmath}
\usepackage{tabularx}
\usepackage{xurl} 
\usepackage[colorlinks=true, allcolors=blue]{hyperref}

\title{Evaluating Electrostatic Embedding MLIP/MM for Relative Binding Free Energy Calculations}

\author{Stephen E. Farr}
\affiliation{Acellera Labs, C Dr Trueta 183, 08005, Barcelona, Spain}
\author{Gianni De Fabritiis}
\email{g.defabritiis@acellera.com}
\affiliation{Computational Science Laboratory, Universitat Pompeu Fabra, Barcelona Biomedical Research Park (PRBB), C Dr. Aiguader 88, 08003, Barcelona, Spain}
\altaffiliation{Acellera, 38350 Fremont Blvd 203 Fremont CA, 94536 USA}
\altaffiliation{Instituci\'o Catalana de Recerca i Estudis Avan\c{c}ats (ICREA), Passeig Lluis Companys 23, 08010 Barcelona, Spain}


\begin{document}

\begin{abstract}
Alchemical relative binding free energy (RBFE) calculations are limited by the
fixed-charge approximation of classical force fields. Hybrid machine learning
interatomic potential/molecular mechanics (MLIP/MM) schemes correct ligand
strain, but under mechanical embedding still describe ligand--environment
electrostatics with static point charges. Electrostatic embedding schemes
coupling machine-learned charges to the MM environment have been proposed and
validated against QM/MM for simple systems, but not tested in a production alchemical workflow. We
take the electrostatic embedding scheme of Semelak et al.\ and evaluate it on
protein--ligand RBFE. We trained a TensorNet2 model, \texttt{AceFF-2-RESP-1}, on
$10^{6}$ conformations from the AceFF dataset, jointly predicting energies,
forces and Restrained Electrostatic Potential (RESP) charges. We chose RESP
over MBIS for commensurability with the AMBER-family force field it couples to.
The predicted charges enter the short-range direct-space part of the particle mesh
Ewald sum, with Thole damping to prevent polarization catastrophes during
alchemical transformations. 
We tested the scheme across five targets from the Wang et al.\ benchmark set,
fixed in advance by a prior study, with three replicates per edge and matched
protocols. Electrostatic embedding improved every accuracy and correlation
metric for TYK2 ($\Delta\Delta G$ RMSE $0.86 \rightarrow 0.45$~kcal/mol against
GAFF2), but performed comparably to the classical and mechanical-embedding
baselines for CDK2, thrombin, p38 and JNK1. Standard single-molecule energy and
charge benchmarks were not good predictors of this target-dependent outcome. TYK2 combined
good $\Delta\Delta G$ accuracy with the lowest force error on the Schrödinger
benchmark, but this pattern did not hold for the other targets. 
\end{abstract}

\section{Introduction}

Accurate prediction of protein--ligand binding affinities is crucial in drug discovery, particularly during the hit-to-lead and lead optimization phases. Alchemical relative binding free energy (RBFE) calculations are an industry-standard technique for this, used to guide molecular design and prioritize compounds for synthesis~\cite{wang2015accurate, ross2023maximal, schindler2020_merck_set, cournia2017_rbfe_review}. However, the accuracy of RBFE calculations is limited by the quality of the underlying force field~\cite{cournia2020_forcefield_limits}. Traditional molecular mechanics (MM) force fields, such as GAFF~\cite{wang2004development, wang2006automatic}, CGenFF~\cite{vanommeslaeghe2010charmm, vanommeslaeghe2012automation}, and OpenFF~\cite{qiu2021development, boothroyd2023development}, rely on fixed-charge parameterizations, typically AM1-BCC~\cite{jakalian2000_am1bcc_I, jakalian2002_am1bcc_II} or RESP~\cite{bayly1993_resp}, combined with a fixed protein force field such as ff14SB~\cite{maier2015ff14sb}. These struggle to capture the electronic properties of diverse drug-like molecules, particularly rare functional groups, out-of-parameter-space torsions, and effects such as electronic polarization and charge transfer~\cite{cisneros2016_polarization_review}. Polarizable force fields such as AMOEBA and the Drude oscillator models~\cite{ponder2010_amoeba, lemkul2016_drude} address part of this, but at substantial parameterization and computational cost.

Machine learning interatomic potentials (MLIPs) offer a compelling alternative to classical force fields~\cite{Behler2007NNP, Bartok2010GAP, Smith2017ANI1, Unke2021MLFFReview, duval2023hitchhiker}. Trained on large quantum mechanical (QM) datasets~\cite{levine2025openmolecules2025omol25, kim_pubchem_2025}, modern MLIPs achieve near-QM accuracy for molecular energies and forces at a fraction of the computational cost of ab initio methods. Architectures now routinely applied to drug-like molecules include ANI-2x~\cite{ANI2x}, MACE and MACE-OFF23~\cite{batatia_mace_2023, kovacs2023mace}, NequIP~\cite{batzner_e3-equivariant_2022}, AIMNet2~\cite{anstine2024aimnet2}, TensorNet~\cite{TensorNet, simeon_broadening_2025}, Egret~\cite{mann_egret-1_2025}, SO3LR~\cite{kabylda_molecular_2025}, FeNNix-Bio1~\cite{fennix_bio1}, and the recent OMol25-trained OrbMol and UMA models~\cite{orbmol_huggingface, wood_uma_2025, levine2025openmolecules2025omol25}. Simulating an entire protein-solvent system at the MLIP level remains too expensive for routine use, so hybrid MLIP/MM (or MLIP/MM) schemes, which apply the same ligand/environment partitioning as classical QM/MM~\cite{warshel1976_qmmm, field1990_qmmm, senn2009_qmmm_review}, have become the standard route for using machine learning in biomolecular simulation~\cite{Rufa2020.07.29.227959, sabanes2024enhancing, sabanes_zariquiey_quantumbind-rbfe_2025}. In these schemes the ligand is modeled with the MLIP, while the surrounding protein and solvent are modeled with a classical MM force field~\cite{eastman2023openmm, galvelis2023nnp}.

In our previous work~\cite{sabanes_zariquiey_quantumbind-rbfe_2025}, building on~\cite{sabanes2024enhancing}, we showed that an MLIP/MM framework for RBFE calculations using the TensorNet architecture~\cite{TensorNet, simeon_broadening_2025} (via the AceFF-1.0 potential~\cite{aceff_huggingface}) improved binding affinity predictions compared to classical force fields, such as GAFF2. That framework used mechanical embedding: the MLIP models intramolecular ligand strain and local bonded configurations, but the non-bonded interactions between the ligand and its environment still use static, pre-computed MM parameters, e.g.\ fixed AM1-BCC charges~\cite{jakalian2000_am1bcc_I}. This accounts for ligand strain but not for the dynamic, conformation-dependent electronic structure changes that occur as the ligand samples different geometries in a polarized binding pocket. Karwounopoulos et al.~\cite{karwounopoulos2025_endstate_mechanical_embedding} recently found no statistically significant RBFE improvement from ML/MM end-state corrections with mechanical embedding across 108 alchemical edges. We note that their work used the end-state correction method with non-equilibrium switching, in contrast to our method~\cite{sabanes_zariquiey_quantumbind-rbfe_2025} which uses the MLIP directly for all intermediate lambda states and uses standard equilibrium FEP. Karwounopoulos et al attributed the lack of improvement to the protein-ligand interaction still being described by unchanged MM parameters. This motivates the present work: if the intermolecular term is the bottleneck, improving the intramolecular description alone cannot help, and the coupling itself has to be upgraded.

Electrostatic embedding addresses this by letting the MLIP region couple continuously to the surrounding MM electrostatic field. Several strategies for this have been proposed, including field-conditioned architectures~\cite{gastegger2021_fieldschnet}, machine-learned multipoles~\cite{thurlemann2022_atomic_multipoles}, and explicit electrostatic embedding of MLIPs in condensed-phase QM/MM~\cite{boselt2021_ml_qmmm, zinovjev2023_electrostatic_embedding_mlp, zinovjev2024_emle_engine}. Grassano et al.~\cite{grassano2024_embedding_schemes} compared mechanical and electrostatic embedding for MLIP/MM and found MBIS charges to agree best with reference QM/MM, and Morado et al.~\cite{morado2025_enhancing_electrostatic_embedding} have recently applied electrostatic MLIP/MM embedding to free energy calculations. A broader family of MLIPs predict charges or long-range electrostatics as part of the energy model, including AIMNet2~\cite{anstine2024aimnet2}, 4G-HDNNP~\cite{ko_fourth-generation_2021}, latent Ewald summation~\cite{cheng_latent_2025}, charge-equilibration approaches~\cite{fennix_bio1}, and the Nutmeg models~\cite{eastman_nutmeg_2024}, which, like the present work, build on TensorNet but inject \emph{precomputed} partial charges rather than predicting them. Most relevant to us, Semelak et al.~\cite{semelak_advancing_2025} treat the ML subsystem as an electrostatic entity that interacts with its MM environment through geometry-dependent, machine-learned partial charges, with added polarization and electronic distortion corrections. They combined energy, force and MBIS charge prediction into a single network, ANI-MBIS, and validated it against reference QM/MM on solvation free energies, conformational landscapes, and small ligand-analogue complexes. What their work does not establish is whether the scheme holds up in a production alchemical workflow: whether the coupling term is numerically stable across a full $\lambda$ schedule, whether it is affordable at the scale of a congeneric-series campaign, and whether the improved physics actually gives better binding affinity predictions.

This paper answers those questions. We take the electrostatic embedding scheme of Semelak et al.~\cite{semelak_advancing_2025} essentially unchanged, and make two practical changes. First, we train the charge model on Restrained Electrostatic Potential (RESP) charges~\cite{bayly1993_resp} rather than MBIS charges~\cite{verstraelen2016_mbis}. RESP is the charge definition the AMBER-family fixed-charge force fields were parameterized against~\cite{cornell1995_amber_ff, wang2004development}, so the predicted charges are commensurable both with the MM environment they couple to and with the AM1-BCC~\cite{jakalian2000_am1bcc_I} baseline they replace. This matters more for an alchemical calculation than for a single-point comparison, since the ligand charges must stay consistent with the surrounding force field at every value of $\lambda$. Second, we build on the TensorNet2 architecture~\cite{TensorNet, simeon_broadening_2025} underlying the AceFF potentials~\cite{aceff_huggingface, farr2026aceff2} instead of ANI, so the same network serves as both the ligand potential and the charge source inside an MLIP/MM stack already validated for RBFE~\cite{sabanes_zariquiey_quantumbind-rbfe_2025}.

The main contribution, however, is the evaluation rather than the model. We ran the scheme through a full alchemical protein--ligand RBFE campaign: five targets from the Wang et al.\ benchmark set~\cite{wang2015accurate, ross2023maximal, schindler2020_merck_set, cournia2017_rbfe_review}, three independent replicates per edge, under the protocol established for QuantumBind-RBFE~\cite{sabanes_zariquiey_quantumbind-rbfe_2025}, so the results are directly comparable to both a classical GAFF2 baseline and a mechanical-embedding AceFF-1.0 baseline on identical systems and edges. To our knowledge, this is the first assessment of electrostatic-embedding MLIP/MM against experimental binding affinities on a standard congeneric-series benchmark. Holding the systems, edges, protocol, and sampling fixed lets us separate the effect of the electrostatic coupling from the effect of the underlying potential, and lets us ask a question that standard single-molecule energy/force benchmarking cannot: whether accuracy on static energy, force, and charge test sets predicts accuracy in the free energies we actually care about.

\section{Methods}
\subsection{Quantum Chemical Dataset and Neural Network Training}

To train a model capable of accurately predicting geometry-dependent electrostatic properties, we randomly sampled a $10^6$ conformation subset from the AceFF dataset~\cite{farr2026aceff2}, which was built from PubChem molecules~\cite{kim_pubchem_2025} at the $\omega$B97M-V/def2-TZVPPD level of theory~\cite{mardirossian__2016}. Restrained Electrostatic Potential (RESP) partial charges~\cite{bayly1993_resp} were calculated at the HF/6-31G(d) level of theory, the standard used in the parameterization of the AMBER-family fixed-charge force fields~\cite{cornell1995_amber_ff, wang2004development}, using PySCF~\cite{sun_recent_2020} and the GPU-accelerated \texttt{gpu4pyscf} package~\cite{li2024introducting, wu2024enhancing}. 

We trained a compact architecture based on the TensorNet2 design~\cite{TensorNet, simeon_broadening_2025}, as implemented in TorchMD-Net~\cite{pelaez2024torchmd}, to simultaneously predict total potential energy $E$, atomic forces $\mathbf{F}$, and conformational-dependent RESP partial charges $q^{\text{RESP}}$. The network was optimized using a composite loss function $\mathcal{L}$:
\begin{equation}
    \mathcal{L} = w_{e} \mathcal{L}_{\text{energy}} + w_{f} \mathcal{L}_{\text{forces}} + w_{q} \mathcal{L}_{\text{charges}}
\end{equation}
where $w_e$, $w_f$, and $w_q$ are weighting factors balancing the relative contributions of energy, force, and charge mean squared errors, respectively. Jointly fitting energetic and electrostatic targets in this way follows earlier multi-task potentials~\cite{unke2019_physnet, thurlemann2022_atomic_multipoles, mace_polar}.

\subsection{MLIP/MM Electrostatic Embedding Scheme}

To couple the machine learning potential with the surrounding classical environment, we adopted an electrostatic embedding framework extending the principles described by Semelak et al.~\cite{semelak_advancing_2025}. The total potential energy of the multiscale system is partitioned as:
\begin{equation}
    E_{\text{MLIP/MM}}(\mathbf{r}) = E_{\text{MLIP}}(\mathbf{r}_{l}) + E_{\text{MM}}(\mathbf{r}_{p+s}) + E^{\text{interaction}}_{\text{MLIP/MM}}(\mathbf{r}_{l}, \mathbf{r}_{p+s})
\end{equation}
where $\mathbf{r}_l$ denotes the ligand coordinates and $\mathbf{r}_{p+s}$ denotes the combined protein and solvent coordinates. Here, $E_{\text{MLIP}}$ is the vacuum energy of the ligand predicted by the MLIP, while $E_{\text{MM}}$ is the energy of the protein and solvent environment computed via the classical MM force field. Both terms match those utilized in traditional mechanical embedding schemes~\cite{Rufa2020.07.29.227959, sabanes2024enhancing, sabanes_zariquiey_quantumbind-rbfe_2025}.

The coupling term $E^{\text{interaction}}_{\text{MLIP/MM}}$ models non-bonded interactions between the MLIP ligand region and the surrounding MM environment:
\begin{equation}
    E^{\text{interaction}}_{\text{MLIP/MM}}(\mathbf{r}_{l}, \mathbf{r}_{p+s}) = E_{\text{LJ}}(\mathbf{r}_l, \mathbf{r}_{p+s}) + E_{\text{coul}}(\mathbf{r}_l, \mathbf{r}_{p+s})
\end{equation}
While the Lennard-Jones (LJ) interactions $E_{\text{LJ}}$ retain fixed classical MM parameters ($\epsilon_i, \sigma_i$), the Coulombic interaction $E_{\text{coul}}$ incorporates conformationally dependent ligand partial charges $q_i(\mathbf{r}_l)$ predicted directly by the MLIP:
\begin{equation}
    \label{eq:coul}
    E_{\text{coul}}(\mathbf{r}_l, \mathbf{r}_{p+s}) = \sum_{i \in l, j \in p+s} \frac{q_i(\mathbf{r}_l) q_j}{4 \pi \varepsilon_0 r_{ij}} + E_{\text{pol}}(\mathbf{r}_l, \mathbf{r}_{p+s}) + E_{\text{distortion}}(\mathbf{r}_l, \mathbf{r}_{p+s})
\end{equation}

The polarization energy term $E_{\text{pol}}$ accounts for the interaction of the external MM electric field $\vec{E}_{\text{MM}}$ acting on the isotropic atomic polarizabilities $\alpha_i$ of the ligand atoms:
\begin{equation} \label{eq:epol}
    E_{\text{pol}}(\mathbf{r}_l, \mathbf{r}_{p+s}) = - \frac{1}{\epsilon} \sum_{i \in l} \alpha_i \left| \vec{E}_{\text{MM}}(\mathbf{r}_i) \right|^2
\end{equation}
Following Semelak et al.~\cite{semelak_advancing_2025}, the electronic distortion energy penalty is fixed at $E_{\text{distortion}} = -0.5 E_{\text{pol}}$, which follows the standard linear-response argument for induction energy~\cite{israelachvili_or_stone_intermolecular_forces}. Atomic polarizabilities $\alpha_i$ are fixed hyperparameters: for the elements it covers, values are taken from the ANI-MBIS table~\cite{semelak_advancing_2025}, with the remaining elements taken from~\cite{vanduijnen1998_atomic_polarizabilities}. The effective dielectric constant $\epsilon$ is initially set to 2.

\subsection{Short-Range PME Modification and Thole Damping}

To facilitate efficient computation during molecular dynamics (MD) simulations under Particle Mesh Ewald (PME) periodic boundary conditions~\cite{darden1993_pme, essmann1995_smooth_pme}, we reformulate the electrostatic coupling. Specifically, the dynamic MLIP charges $q_i(\mathbf{r}_l)$ are evaluated only within the short-range, direct-space contribution of the PME sum:
\begin{equation} \label{eq:coul_sr}
    E^{\text{SR}}_{\text{coul}} = \sum_{i \in l, j \in p+s} \frac{q_i(\mathbf{r}_l) q_j}{4 \pi \varepsilon_0 r_{ij}} \operatorname{erfc}(\beta r_{ij})
\end{equation}
In practice, the full-system MM Coulomb energy is computed as usual using fixed baseline MM charges. We then calculate the short-range contribution difference between the dynamic MLIP charges and static MM charges, adding this $\Delta E^{\text{SR}}_{\text{coul}}$ as an additive correction to the total potential energy. This avoids backpropagating gradients through the reciprocal-space Ewald mesh, reducing computational overhead while keeping an exact short-range edge cutoff. 
Because long-range electrostatics vary negligibly between RESP and AM1-BCC parameterizations, we hypothesize this error can be ignored.

Furthermore, when evaluating the MM electric field acting on ligand atom $i$, we introduce a Thole damping function $\lambda_{\text{thole}}(r_{ij})$~\cite{thole1981_damping, vanduijnen1998_atomic_polarizabilities}. While optional in standard equilibrium MD, damping is critical within the Alchemical Transfer Method (ATM) framework~\cite{azimi2022_atm, wu2021_atm_original} to prevent polarization catastrophes (energy divergence) when alchemical transformations force atoms into unphysically close proximity. The damped electric field $\vec{E}_i^{\text{damped}}$ evaluated from the gradient of Eq.~\ref{eq:coul_sr} is expressed as:
\begin{equation}
    \vec{E}_i^{\text{damped}} = \sum_{j \in p+s} q_j \lambda_{\text{thole}}(r_{ij}) \left[ \frac{\operatorname{erfc}(\beta r_{ij})}{r^2_{ij}} + \frac{2 \beta}{\sqrt{\pi}} \frac{e^{-\beta^2 r^2_{ij}}}{r_{ij}} \right] \hat{\vec{r}}_{ij}
\end{equation}
where the Thole damping factor $\lambda_{\text{thole}}(r_{ij})$ is inspired by the linear Thole model~\cite{thole1981_damping}, scaled by the polarizability $\alpha_i$ of the ML atom alone:
\begin{equation} \label{eq:thole_func}
    \lambda_{\text{thole}}(r_{ij}) = 1 - \left(1 + a\,u_{ij} + \tfrac{1}{2}(a\,u_{ij})^2\right) e^{-a\,u_{ij}}, \qquad u_{ij} = \frac{r_{ij}}{\alpha_i^{1/3}}
\end{equation}
with $a$ a dimensionless damping exponent set to $1.3$. This damping term impacts only the alchemical intermediate states; it was introduced because standard softcore and softplus perturbations in the Alchemical Transfer Method are insufficient to prevent $1/r^2$ terms from generating overflowing forces (NaNs) during autograd differentiation. At physical endstates, its physical contribution is negligible, hence the undamped polarization term was suitable in~\cite{semelak_advancing_2025}.

\subsection{Implementation and Alchemical Molecular Dynamics}
\label{sec:implementation}
We implemented this framework within our Alchemical Transfer Method package and make it available to the community. See the Data and Software availability section. The MLIP predicted charges are fed into equation~\ref{eq:coul} and PyTorch Autograd is used to compute the forces on the MLIP and MM atoms. Otherwise it is the same as the mechanical embedding implementation introduced in~\cite{sabanes_zariquiey_quantumbind-rbfe_2025}.

\section{Results}

\subsection{Neural Network Architecture and Joint Learning Performance}
\label{sec:joint_learning}

To jointly model potential energy, interatomic forces, and geometry-dependent partial charges, we implemented a customized TensorNet2 architecture~\cite{TensorNet, simeon_broadening_2025} featuring 2 interaction layers and an embedding dimension of 128. The complete code implementation and model weights are publicly accessible on GitHub.

This is consistent with previous multi-objective electrostatic MLIP architectures, such as MACE-POLAR~\cite{mace_polar} and other joint energy/charge models~\cite{unke2019_physnet, thurlemann2022_atomic_multipoles}: adding an auxiliary charge-prediction loss creates a Pareto trade-off that slightly degrades force accuracy compared to single-task energy/force models. Our joint model reached a final $L_1$ loss of $0.04\,e$ on predicted RESP partial charges and $0.08\,\text{eV}/\text{\AA}$ on atomic forces. For context, the single-task AceFF-2 potential, trained only on energies and forces, reaches a force $L_1$ loss of $0.05\,\text{eV}/\text{\AA}$ on the same benchmark distribution~\cite{farr2026aceff2}.

To check whether this small loss in force accuracy is worth the physical fidelity gained from electrostatic coupling, we evaluated our model on the standard benchmark suite from~\cite{farr2026aceff2}.

\subsection{Evaluation on Strained Conformations: Wiggle150 Benchmark}

We evaluated model stability and energy predictions under extreme conformational strain using the Wiggle150 benchmark~\cite{brew_wiggle150_2025}, which contains 150 highly strained, out-of-equilibrium conformations of three drug-like molecules: adenosine, benzylpenicillin, and efavirenz, with reference energies from DLPNO-CCSD(T)/CBS. We predicted relative energies (against the minimized geometry for each molecule) for all 150 conformations and computed the mean absolute error (MAE) and root mean square error (RMSE) against the reference.

Table~\ref{tab:wiggle} summarizes the results. AceFF-2-RESP-1 outperforms the ANI-2x baseline~\cite{ANI2x} and is within 0.3 kcal/mol MAE of AIMNet2~\cite{anstine2024aimnet2}, but trails the single-task AceFF-2~\cite{farr2026aceff2} by close to 50\% in both MAE and RMSE. OrbMol~\cite{orbmol_huggingface}, with benchmark values taken from~\cite{farr2026aceff2}, is the strongest performer on this benchmark by a wide margin. Neither AceFF-2 nor OrbMol carries the extra burden of jointly predicting partial charges, consistent with the force-accuracy trade-off discussed previously.

\begin{table}[t]
    \centering
    \begin{tabular}{@{}lcc@{}}
        \toprule
        \textbf{Method} & \textbf{MAE (kcal/mol)} & \textbf{RMSE (kcal/mol)} \\ 
        \midrule
        ANI-2X   & 4.41 & 5.41 \\ 
        AceFF-2-resp-1 & 2.67 & 3.41 \\
        AIMNet2  & 2.39 & 3.13 \\ 
        AceFF-2 & 1.76  &  2.34 \\ 
        OrbMol & 0.89 & 1.22 \\
        \bottomrule
    \end{tabular}
    \caption{Wiggle150 benchmark~\cite{brew_wiggle150_2025} results. Reference values for ANI-2x and AIMNet2 are taken from~\cite{brew_wiggle150_2025}; AceFF-2 and OrbMol values from~\cite{farr2026aceff2}.}
    \label{tab:wiggle}
\end{table}

\subsection{Torsional Energy Profiles: Sellers Benchmark}

We evaluated torsional accuracy using the Sellers et al.\ benchmark~\cite{sellers2017comparison}, which contains relaxed torsion scans across 62 drug-like molecules, with CCSD(T)/CBS as the reference. We performed relaxed scans with constrained torsion optimizations using GeomeTRIC~\cite{wang2016geometry}, following the protocol of~\cite{smith_approaching_2019, farr2026aceff2}.

\begin{figure}
    \centering
    \includegraphics[width=0.5\linewidth]{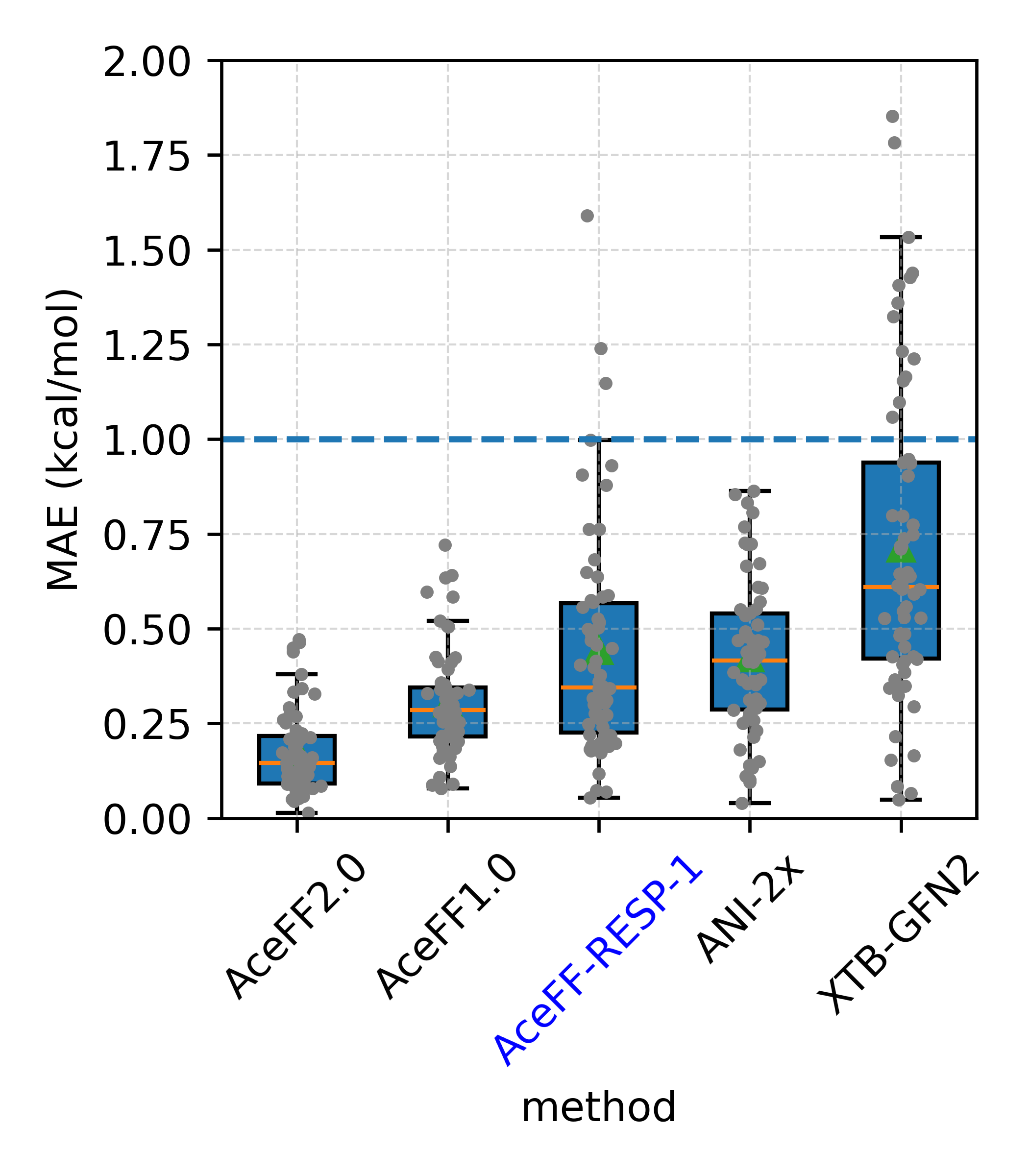}
    \caption{Sellers et al.\ torsion scan benchmark~\cite{sellers2017comparison}. The orange lines are
the median value. The methods are ordered from left to right by median MAE compared
to the coupled cluster reference data. We performed relaxed torsion scans as done in~\cite{smith_approaching_2019, farr2026aceff2}.}
    \label{fig:seller}
\end{figure}

Figure~\ref{fig:seller} shows that our joint energy-and-charge model has lower overall torsional accuracy than the dedicated single-task potentials AceFF-1.0~\cite{aceff_huggingface} and AceFF-2~\cite{farr2026aceff2}. It is roughly comparable to ANI-2x~\cite{ANI2x}, with a lower median error across scanned dihedrals but slightly larger outlier errors, and is substantially more accurate than semi-empirical methods such as GFN2-xTB~\cite{bannwarth_gfn2-xtbaccurate_2019}.

\subsection{Out-of-Distribution Generalizability: Schrödinger Ligand Benchmark}

To assess generalization to complex, drug-like chemical space, we evaluated our model on the Schrödinger ligand benchmark~\cite{farr2026aceff2}, which contains 650 molecular configurations curated from established industry sets (the JACS, Merck, and charge-annihilation subsets)~\cite{ross2023maximal, wang2015accurate, schindler2020_merck_set, farr2026aceff2}. All conformers were relabeled using $\omega\text{B97M-V/def2-TZVPPD}$~\cite{mardirossian__2016} DFT with GPU4PySCF~\cite{sun_recent_2020, li2024introducting, wu2024enhancing}, giving reference potential energies, forces, and RESP partial charges~\cite{bayly1993_resp}.

\begin{figure}[htbp]
    \centering
    \begin{subfigure}[b]{0.48\textwidth}
        \centering
        \includegraphics[width=\linewidth]{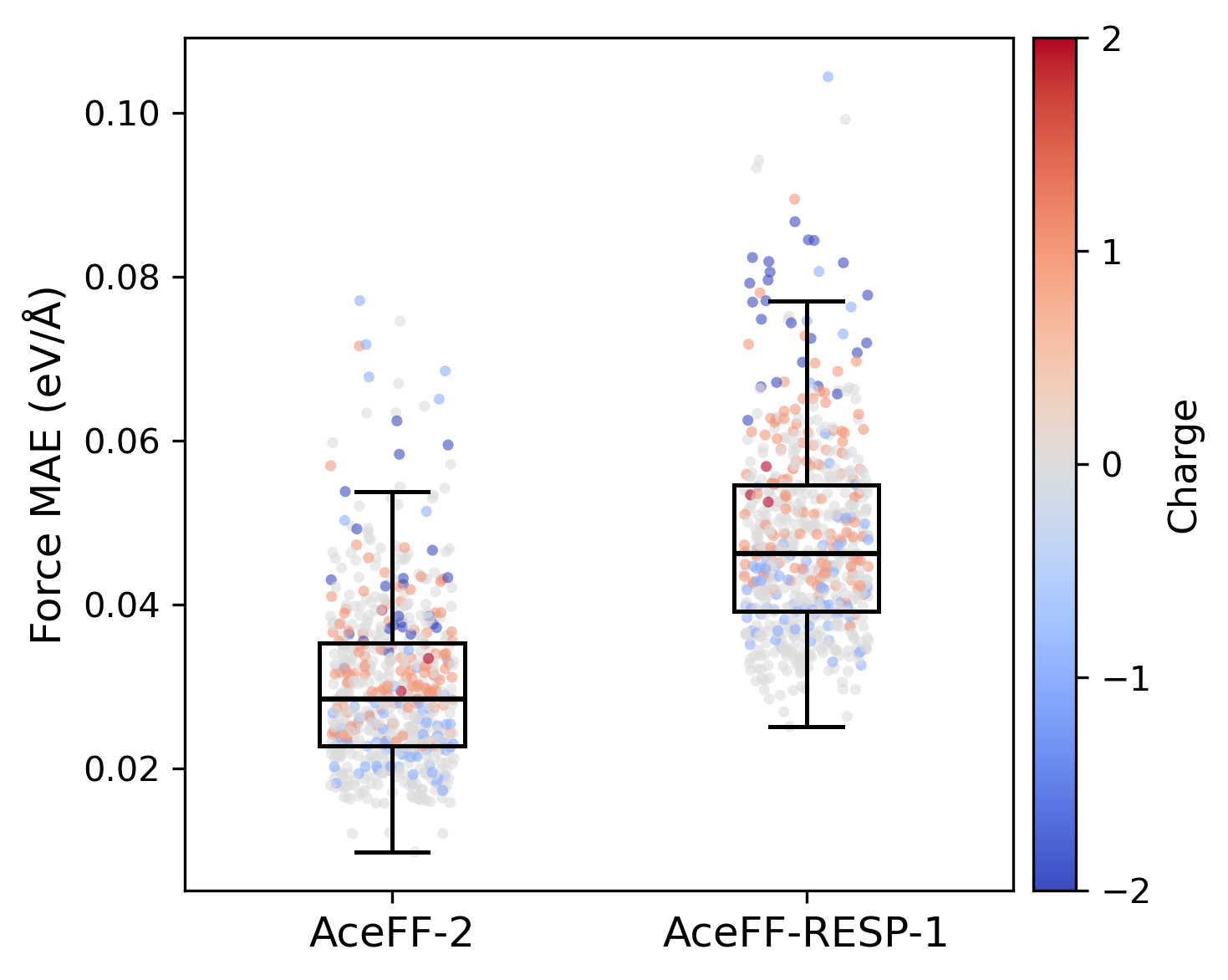}
        \caption{Force MAE distribution relative to AceFF-2.}
        \label{fig:schrodinger_f_mae}
    \end{subfigure}
    \hfill
    \begin{subfigure}[b]{0.48\textwidth}
        \centering
        \includegraphics[width=0.8\linewidth]{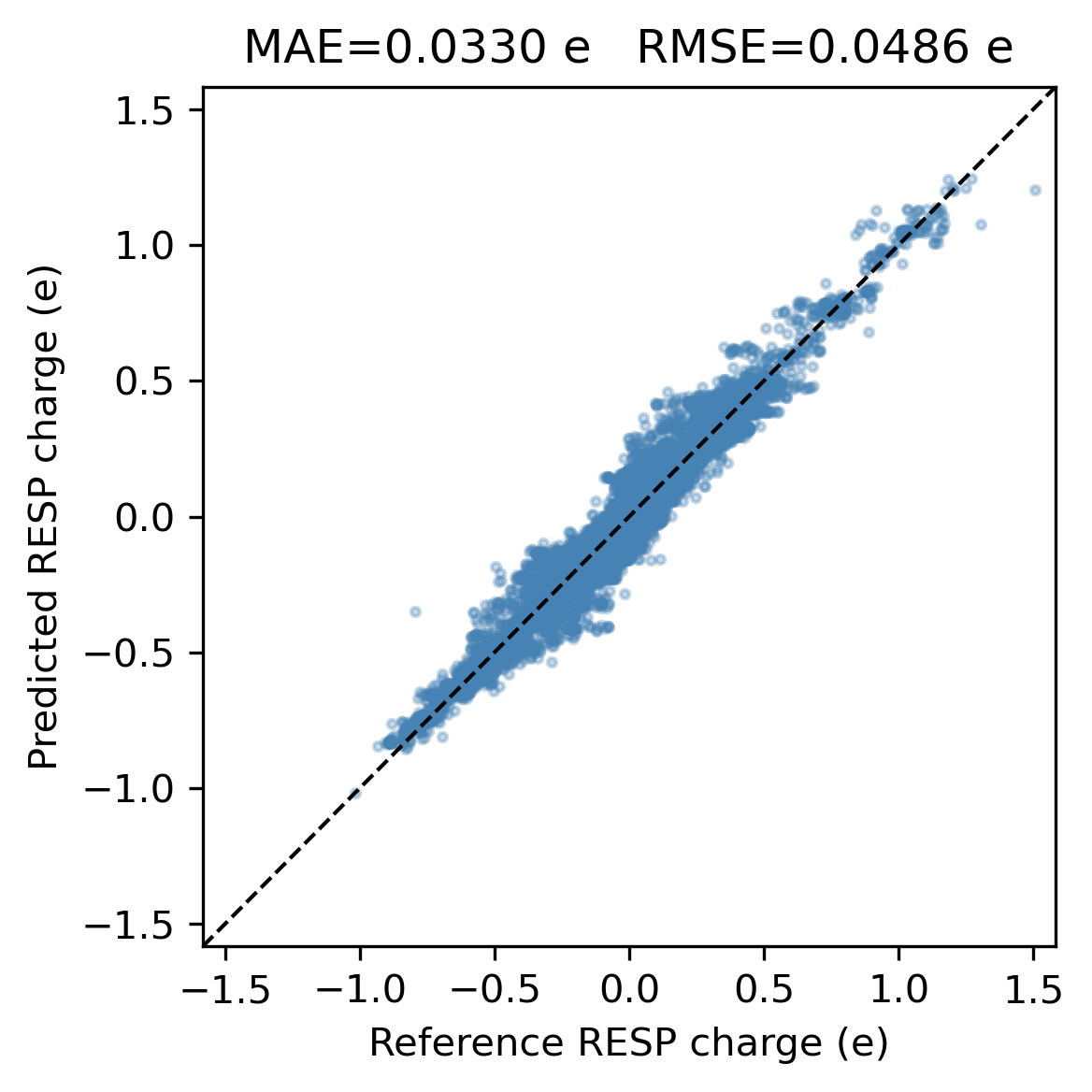}
        \caption{Predicted vs. reference RESP charges.}
        \label{fig:schrodinger_q_correlation}
    \end{subfigure}
    
    \caption{Multi-objective model evaluation on the Schrödinger ligand benchmark (650 conformers). (a) Comparative force error distributions between AceFF-2-RESP-1 and the single-task AceFF-2 reference. (b) Parity plot showing correlation between MLIP-predicted, conformation-dependent partial charges and reference $\omega\text{B97M-V/def2-TZVPPD}$ RESP charges.}
    \label{fig:schrodinger_combined}
\end{figure}

Figure~\ref{fig:schrodinger_combined} shows the multi-objective fidelity of our model on this benchmark. Panel (a) compares the force error distribution of AceFF-2-RESP-1 directly against the single-task AceFF-2 potential~\cite{farr2026aceff2}. Despite the multi-task trade-off from charge fitting, AceFF-2-RESP-1's force accuracy closely tracks AceFF-2.

Panel (b) shows a tight, linear correlation between the MLIP-predicted, geometry-dependent partial charges and the reference RESP charges across the dataset. 
This agreement indicates that the model's charge predictions are transferable to larger unseen ligands.

\subsection{Relative Binding Free Energy (RBFE) Benchmarks}

We benchmarked the \texttt{AceFF-2-RESP-1} potential across five standard protein targets, TYK2, CDK2, Thrombin, p38, and JNK1, from the Wang et al.\ benchmark set~\cite{wang2015accurate}, as curated in the Schrödinger public binding free energy benchmark~\cite{ross2023maximal}. These five targets and their alchemical edges were fixed in advance by the prior QuantumBind-RBFE study~\cite{sabanes_zariquiey_quantumbind-rbfe_2025}, whose systems and protocol we reuse unchanged; they were not selected based on the electrostatic-embedding results reported below.

\begin{figure}[!htbp]
\centering
\begin{subfigure}[t]{0.4\textwidth}
  \includegraphics[width=\linewidth]{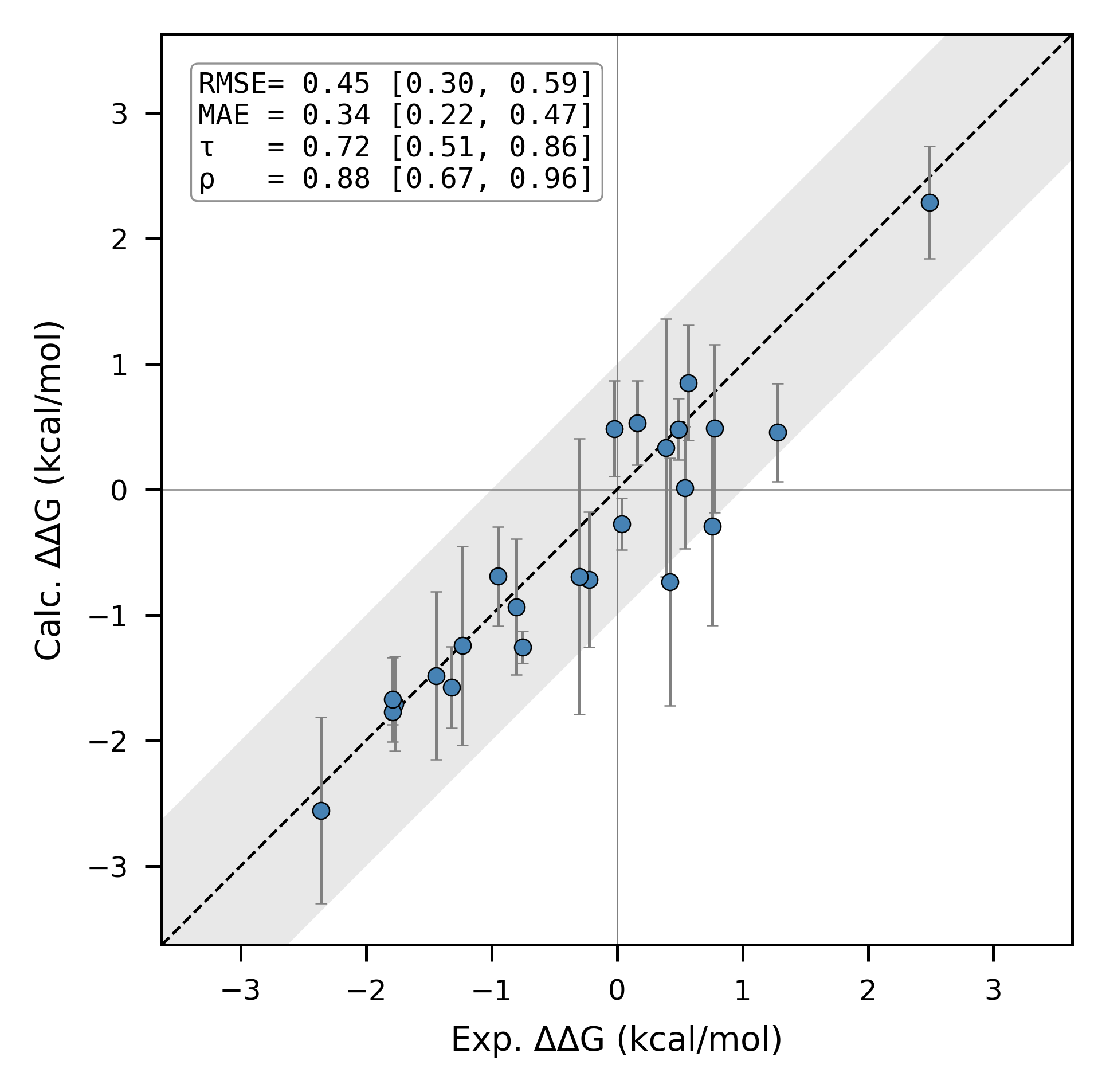}
  \caption{TYK2}
\end{subfigure}\hfill
\begin{subfigure}[t]{0.4\textwidth}
  \includegraphics[width=\linewidth]{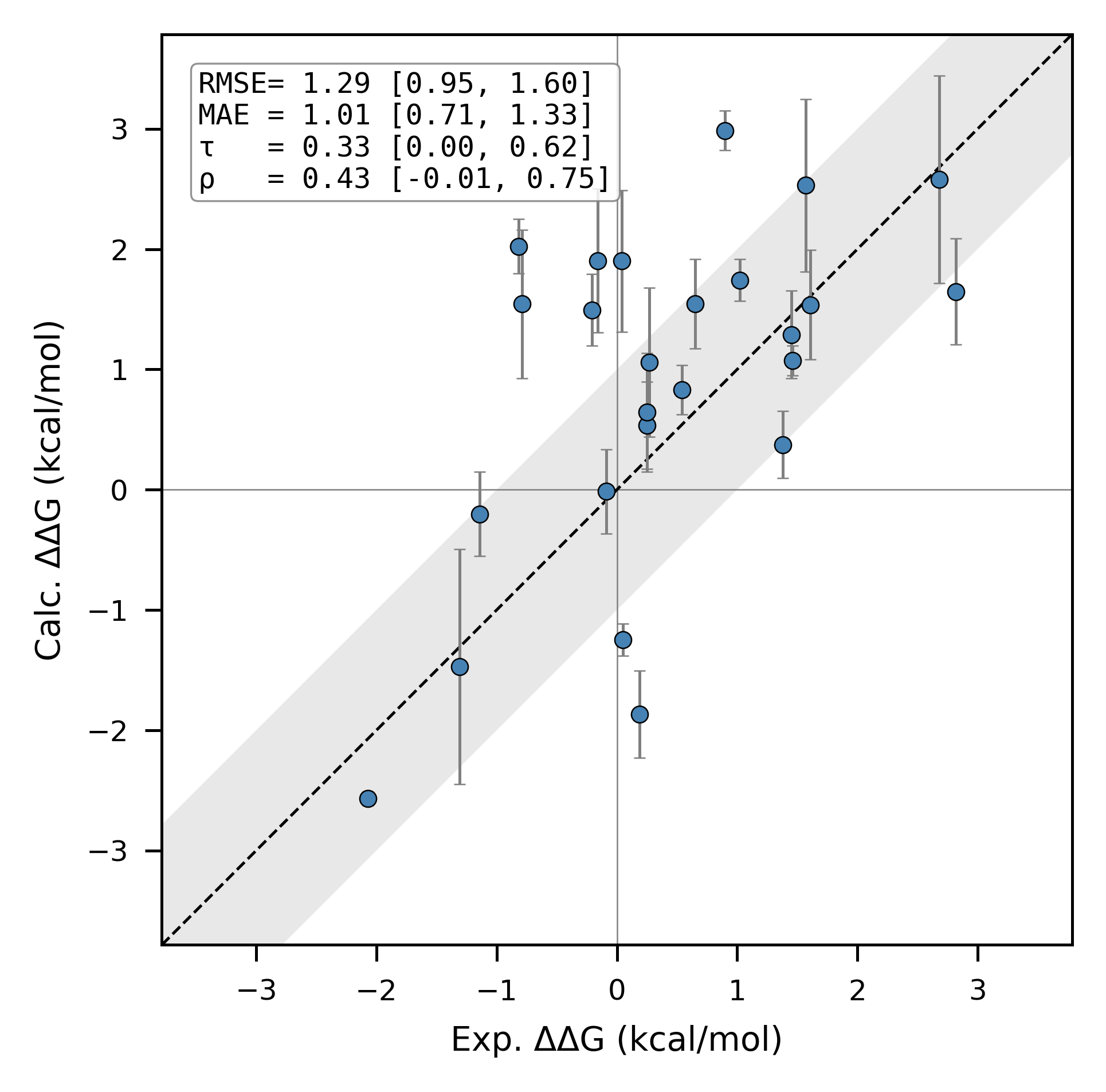}
  \caption{CDK2}
\end{subfigure}

\vspace{4pt}

\begin{subfigure}[t]{0.4\textwidth}
  \includegraphics[width=\linewidth]{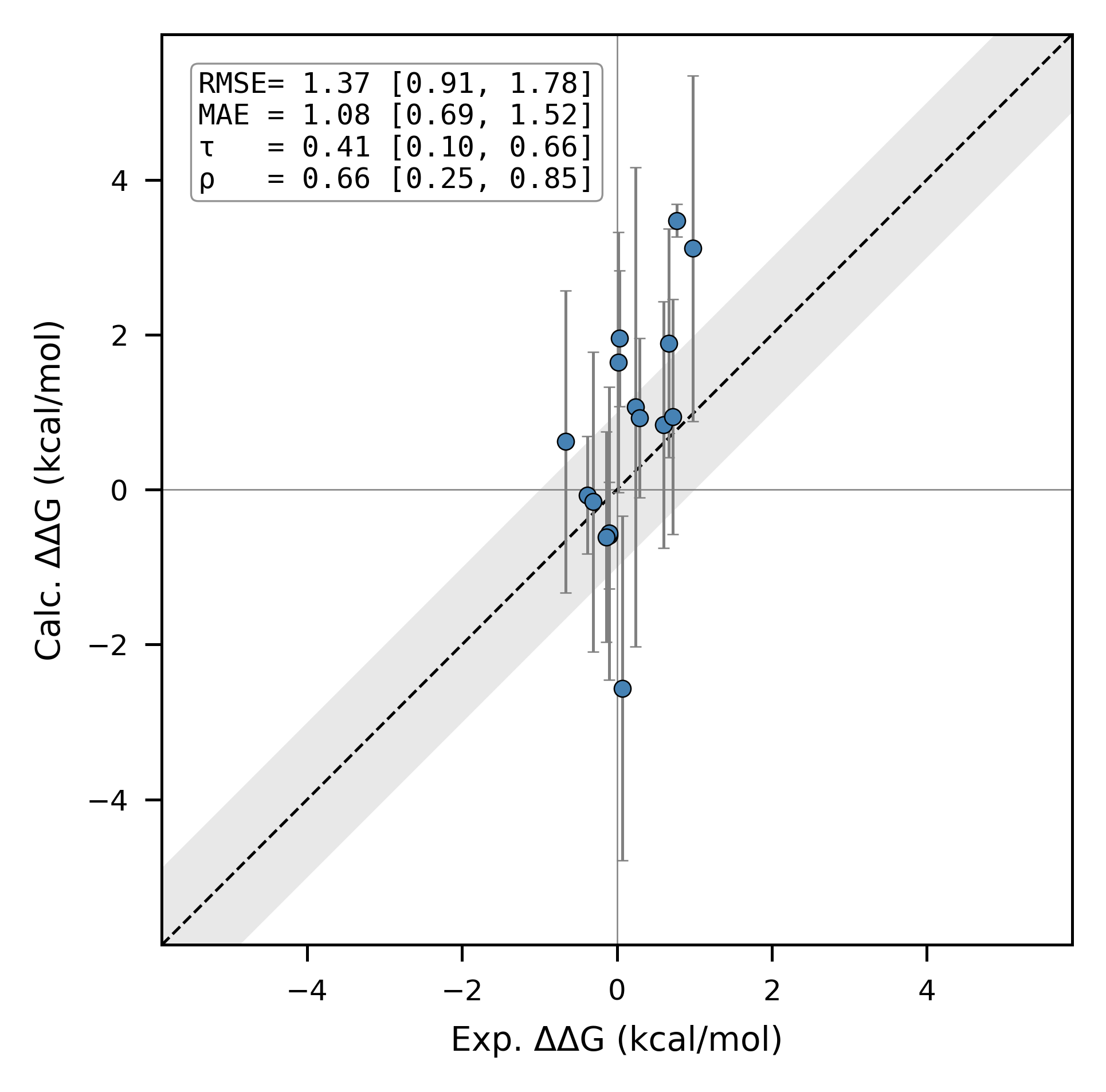}
  \caption{Thrombin}
\end{subfigure}\hfill
\begin{subfigure}[t]{0.4\textwidth}
  \includegraphics[width=\linewidth]{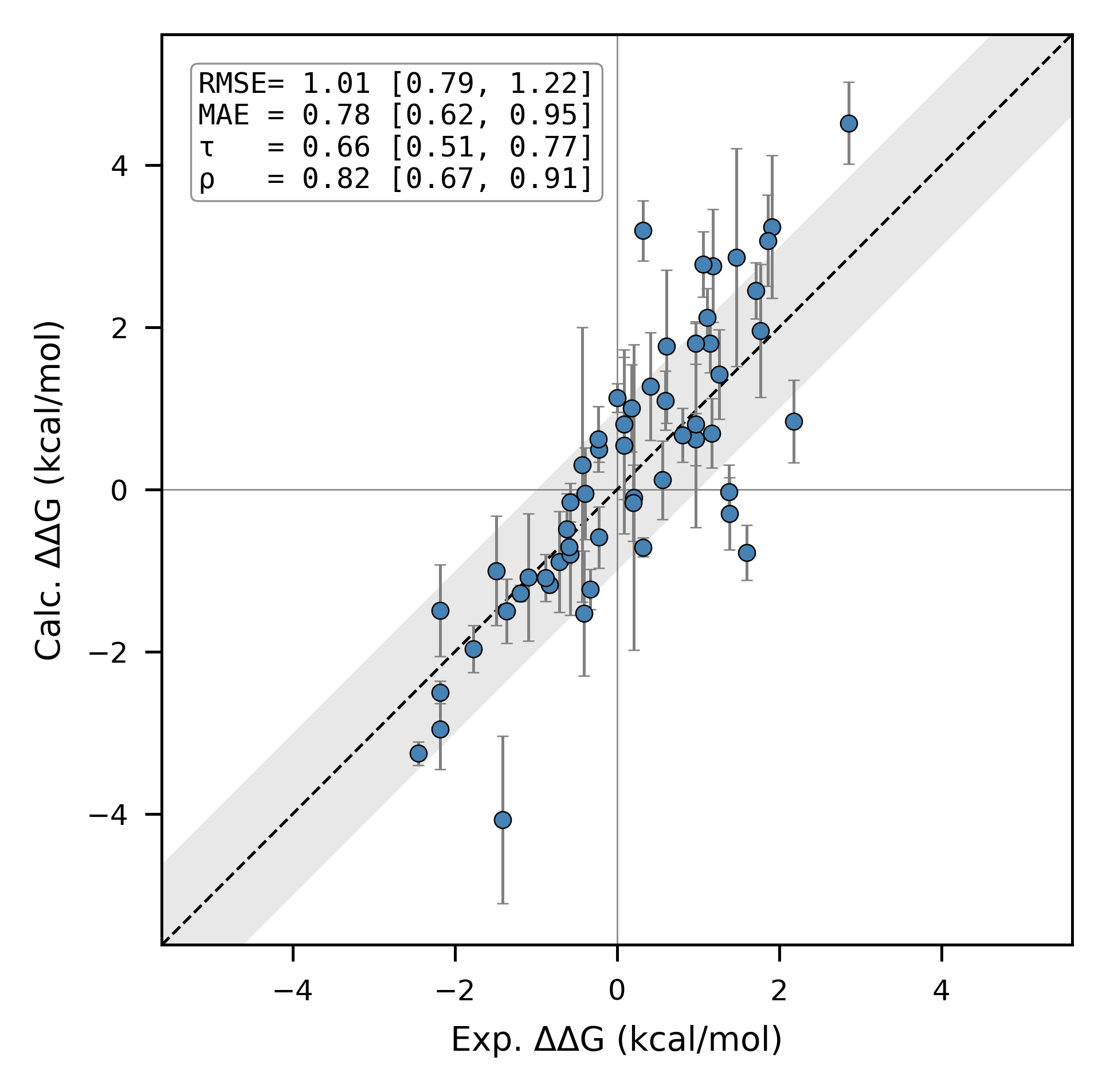}
  \caption{p38}
\end{subfigure}

\vspace{4pt}

\begin{subfigure}[t]{0.4\textwidth}
  \includegraphics[width=\linewidth]{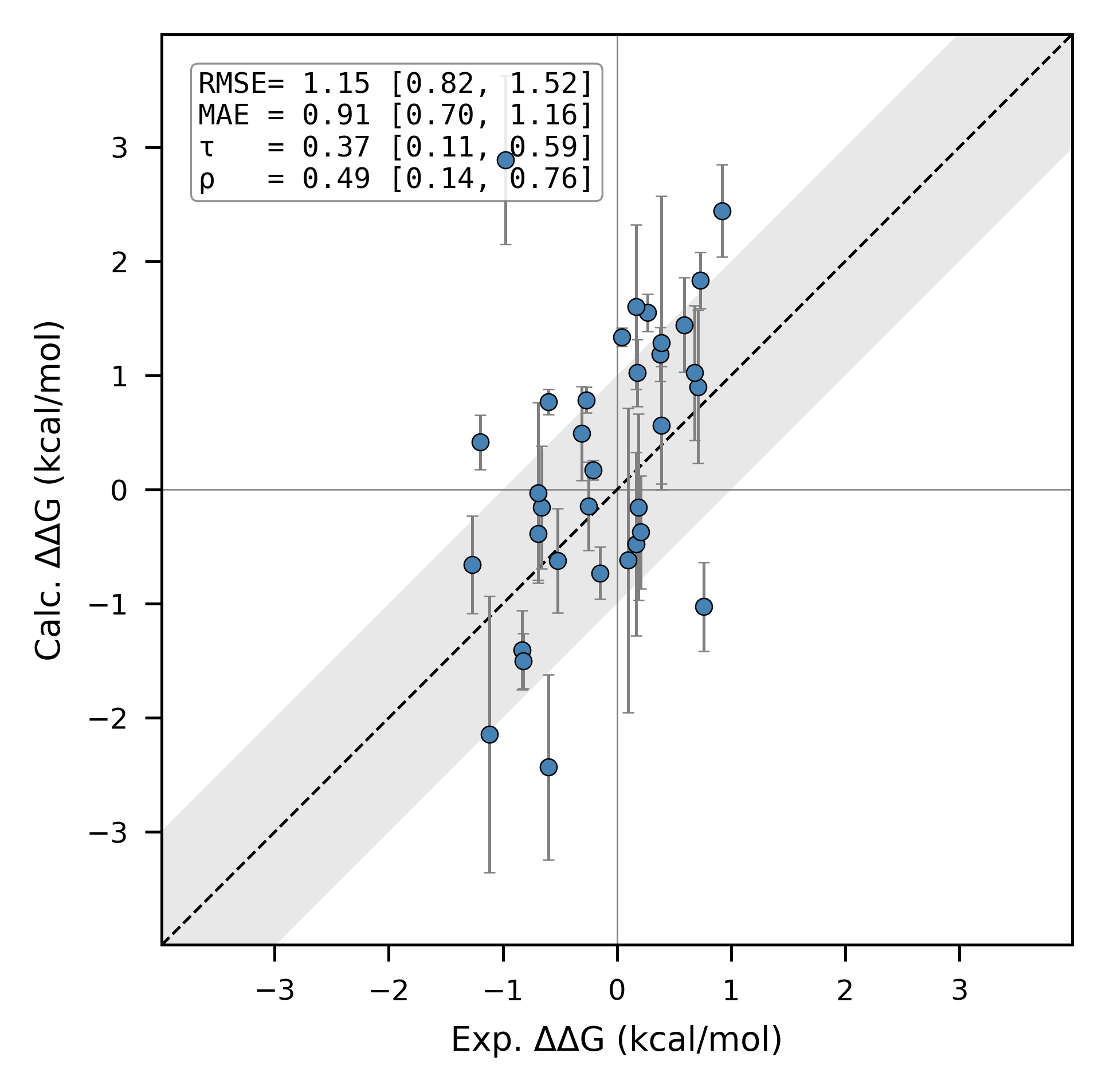}
  \caption{JNK1}
\end{subfigure}\hfill
\begin{subfigure}[t]{0.4\textwidth}\centering
  \vspace{1cm}
\end{subfigure}

\caption{Calculated vs experimental relative binding free energies
($\Delta\Delta G$) on alchemical edges. Each marker shows the mean from the three repeats and the error bars are the standard deviation. 
RMSE, MAE, Kendall $\tau$ and Spearman $\rho$ metrics are annotated in the top-left of each
panel.}
\label{fig:ddg}
\end{figure}

\begin{figure}[!htbp]
\centering
\begin{subfigure}[t]{0.4\textwidth}
  \includegraphics[width=\linewidth]{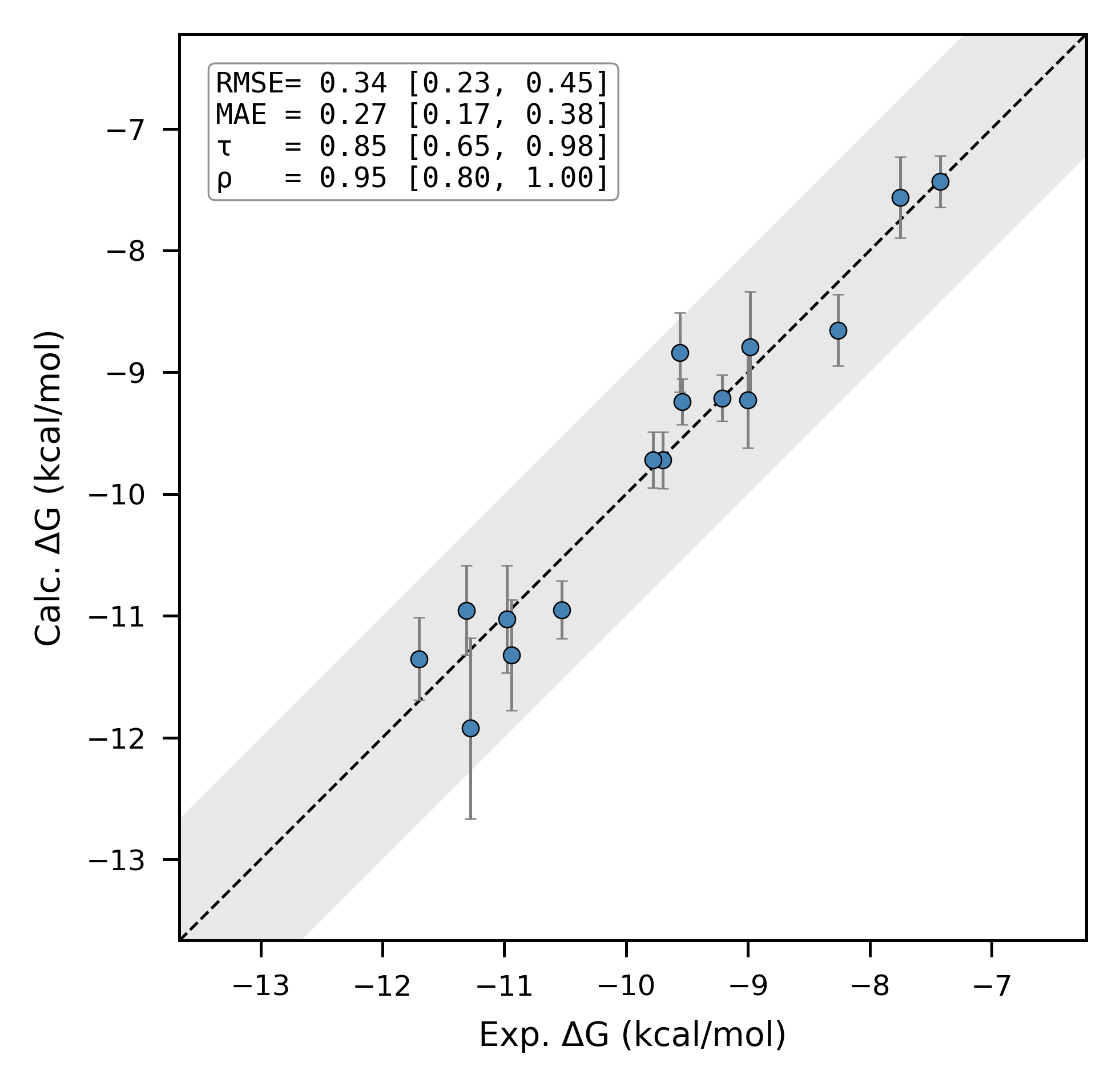}
  \caption{TYK2}
\end{subfigure}\hfill
\begin{subfigure}[t]{0.4\textwidth}
  \includegraphics[width=\linewidth]{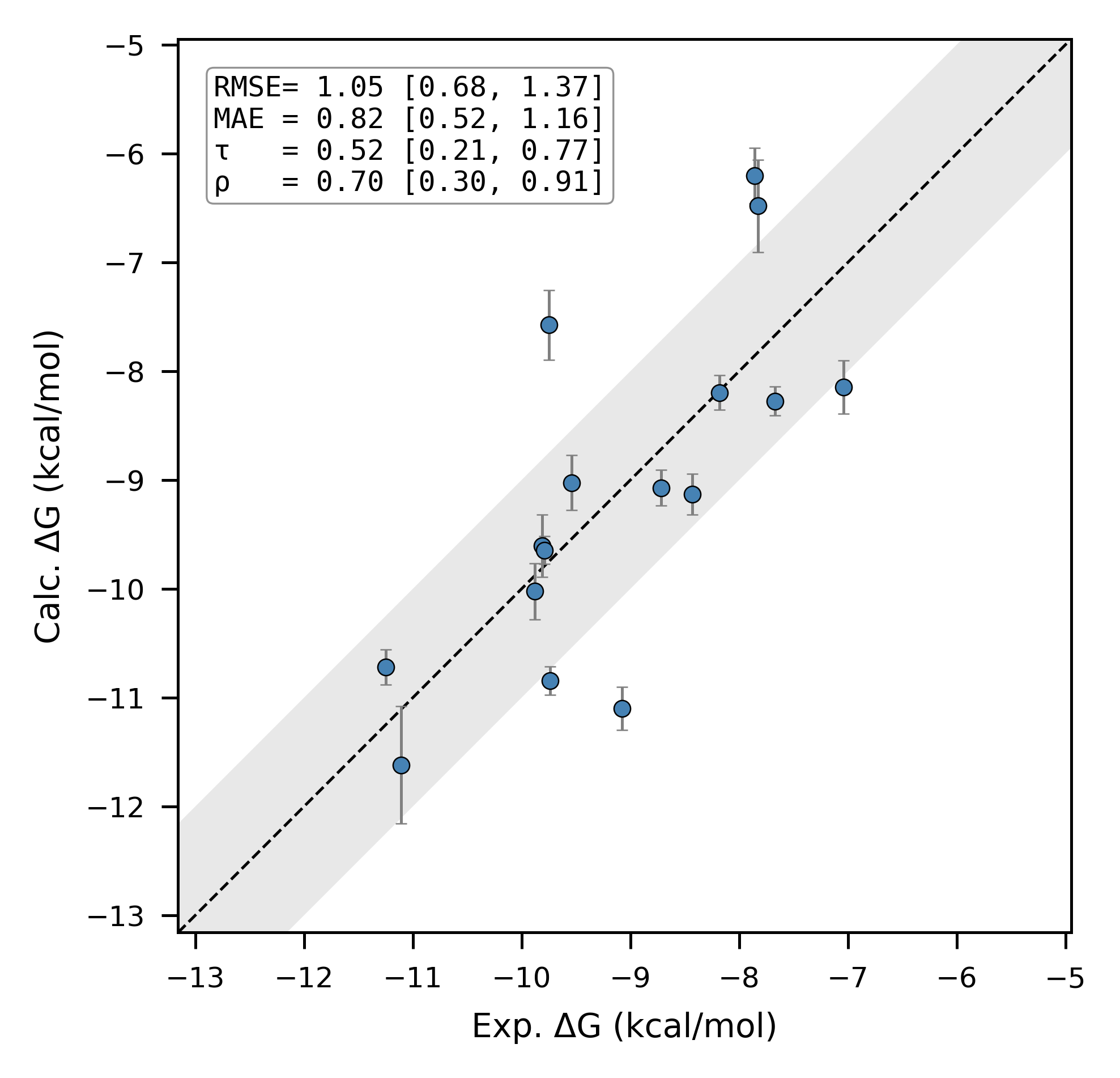}
  \caption{CDK2}
\end{subfigure}

\vspace{4pt}

\begin{subfigure}[t]{0.4\textwidth}
  \includegraphics[width=\linewidth]{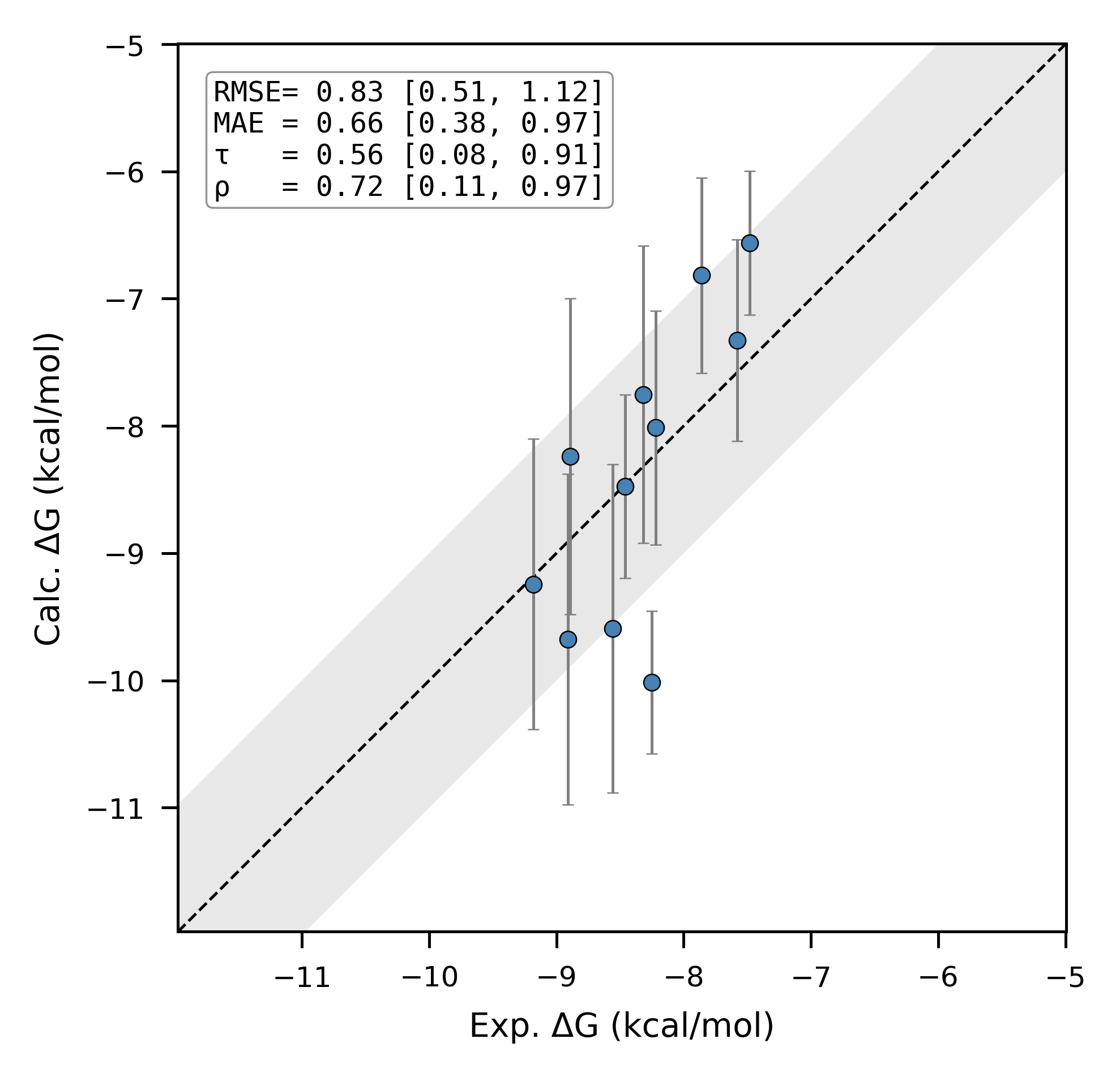}
  \caption{Thrombin}
\end{subfigure}\hfill
\begin{subfigure}[t]{0.4\textwidth}
  \includegraphics[width=\linewidth]{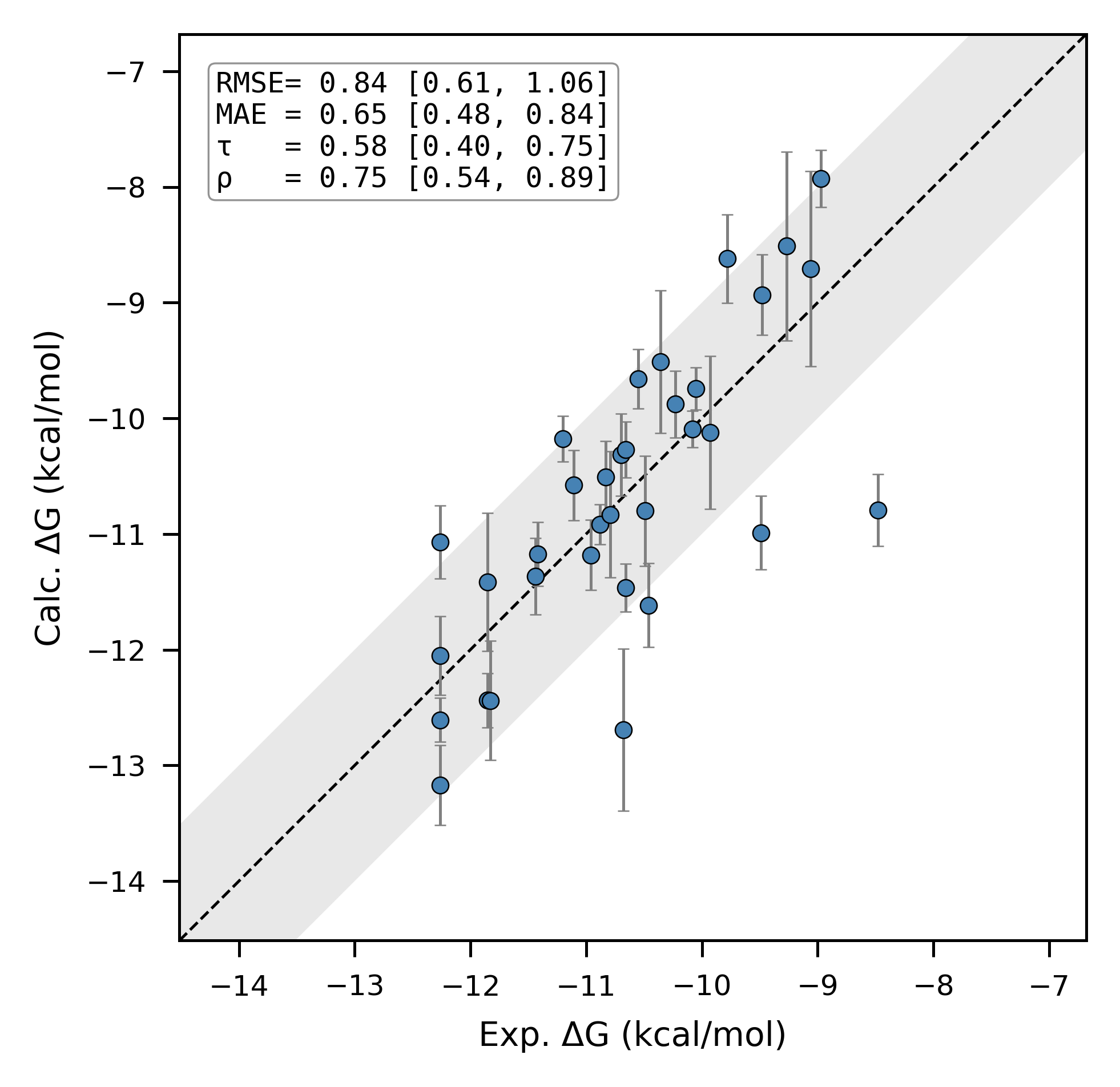}
  \caption{p38}
\end{subfigure}

\vspace{4pt}

\begin{subfigure}[t]{0.4\textwidth}
  \includegraphics[width=\linewidth]{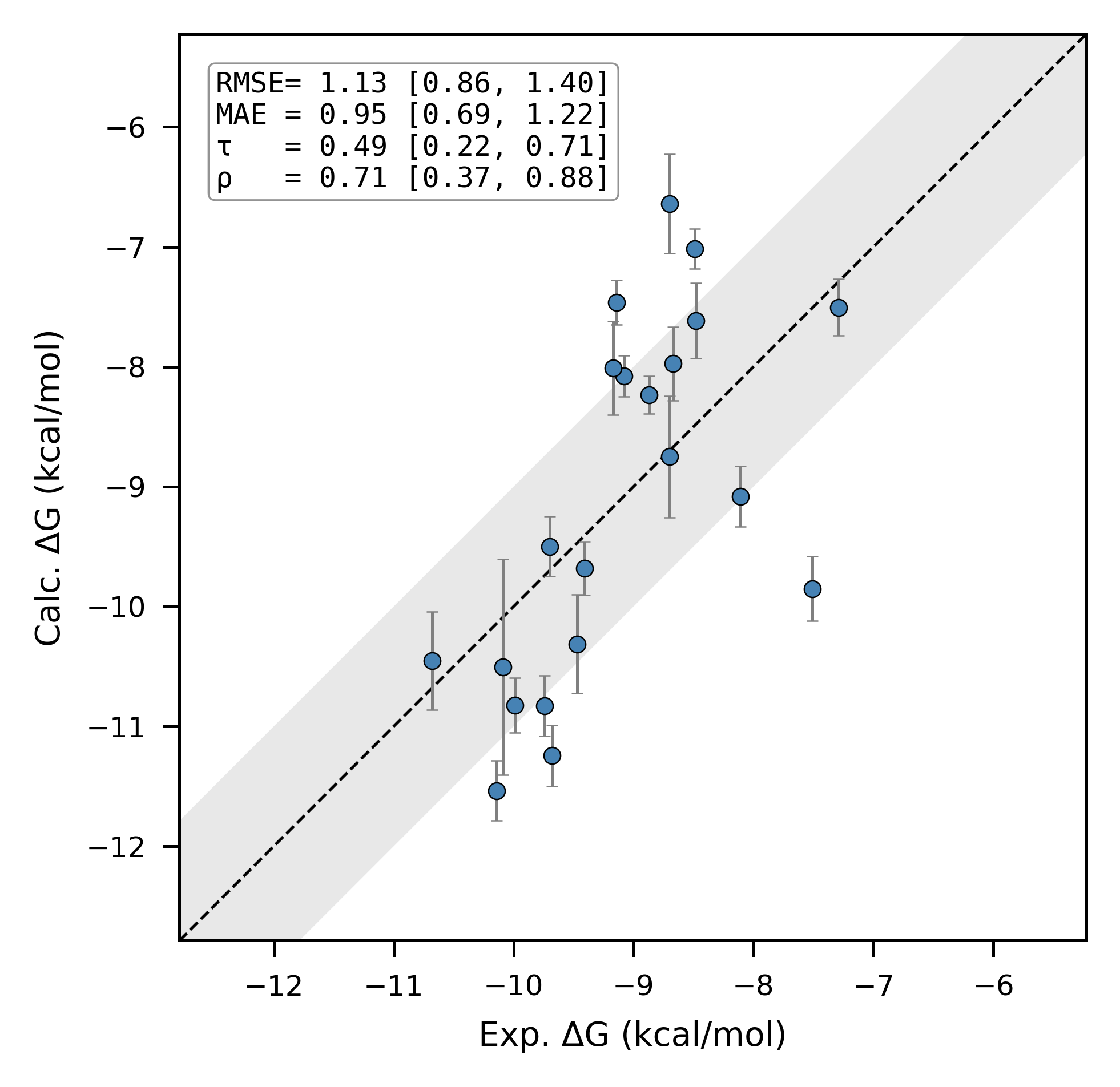}
  \caption{JNK1}
\end{subfigure}\hfill
\begin{subfigure}[t]{0.4\textwidth}\centering
  \vspace{1cm}
\end{subfigure}

\caption{Calculated vs experimental absolute binding free energies
($\Delta G$). Each marker shows the mean from the three repeats and the error bars are the standard deviation. 
RMSE, MAE, Kendall $\tau$ and Spearman $\rho$ metrics are annotated in the top-left of each
panel.}
\label{fig:dg}
\end{figure}

For each target, we report pairwise relative binding free energies ($\Delta\Delta G$) across the alchemical transformation graph (Figure~\ref{fig:ddg}) and absolute binding free energies ($\Delta G$) recovered per ligand by Maximum Likelihood Estimation (MLE) using \texttt{cinnabar}~\cite{Hahnetal2022} (Figure~\ref{fig:dg}). Free energies per $\lambda$-window were estimated with UWHAM/MBAR~\cite{shirts2008_mbar, tan2012_uwham}. Accuracy and correlation metrics against experiment, compared with classical GAFF2 and mechanical-embedding AceFF-1.0, are summarized in Table~\ref{tab:comp_ddg} ($\Delta\Delta G$) and Table~\ref{tab:comp_dg} ($\Delta G$).

All alchemical simulations followed the protocol established by QuantumBind-RBFE~\cite{sabanes_zariquiey_quantumbind-rbfe_2025}. For all five targets, we ran three independent replicates per alchemical edge, with $3\text{ ns}$ of sampling per $\lambda$-window. All other simulation settings, system preparation, soft-core parameters, and thermodynamic analyses followed~\cite{sabanes_zariquiey_quantumbind-rbfe_2025}.

\begin{table}
\caption{Accuracy and correlation metrics for $\Delta\Delta G$ against experiment, compared with GAFF2 and mechanical-embedding AceFF-1.0. Sub/superscripts are bootstrap confidence intervals; best value per system in bold.}
\label{tab:comp_ddg}
\begin{tabular}{llrrrr}
\toprule
& & \multicolumn{4}{c}{$\Delta\Delta G$ (kcal/mol)} \\
\cmidrule(lr){3-6}
System & Model & RMSE & MAE & $\tau$ & $\rho$ \\
\midrule
TYK2 & AceFF-2-RESP-1 & \textbf{0.45}\,$^{0.59}_{0.30}$ & \textbf{0.34}\,$^{0.47}_{0.22}$ & \textbf{0.72}\,$^{0.86}_{0.51}$ & \textbf{0.88}\,$^{0.96}_{0.67}$ \\
 & GAFF2 & 0.86\,$^{1.13}_{0.57}$ & 0.64\,$^{0.89}_{0.42}$ & 0.54\,$^{0.77}_{0.26}$ & 0.69\,$^{0.91}_{0.35}$ \\
 & AceFF1.0 & 0.77\,$^{0.96}_{0.55}$ & 0.57\,$^{0.78}_{0.36}$ & 0.64\,$^{0.81}_{0.44}$ & 0.83\,$^{0.92}_{0.63}$ \\
\midrule
CDK2 & AceFF-2-RESP-1 & \textbf{1.29}\,$^{1.60}_{0.95}$ & 1.01\,$^{1.33}_{0.71}$ & 0.33\,$^{0.62}_{0.00}$ & 0.43\,$^{0.75}_{-0.01}$ \\
 & GAFF2 & 1.50\,$^{1.83}_{1.14}$ & 1.21\,$^{1.57}_{0.87}$ & 0.20\,$^{0.48}_{-0.09}$ & 0.25\,$^{0.60}_{-0.16}$ \\
 & AceFF1.0 & 1.32\,$^{1.74}_{0.83}$ & \textbf{0.88}\,$^{1.29}_{0.52}$ & \textbf{0.40}\,$^{0.70}_{0.03}$ & \textbf{0.50}\,$^{0.82}_{0.02}$ \\
\midrule
Thrombin & AceFF-2-RESP-1 & 1.37\,$^{1.78}_{0.91}$ & 1.08\,$^{1.52}_{0.69}$ & \textbf{0.41}\,$^{0.66}_{0.10}$ & \textbf{0.66}\,$^{0.85}_{0.25}$ \\
 & GAFF2 & 1.42\,$^{1.83}_{0.93}$ & 1.12\,$^{1.56}_{0.71}$ & \textbf{0.41}\,$^{0.69}_{0.11}$ & 0.58\,$^{0.83}_{0.14}$ \\
 & AceFF1.0 & \textbf{1.36}\,$^{1.90}_{0.76}$ & \textbf{1.00}\,$^{1.50}_{0.59}$ & 0.18\,$^{0.52}_{-0.22}$ & 0.25\,$^{0.67}_{-0.31}$ \\
\midrule
p38 & AceFF-2-RESP-1 & \textbf{1.01}\,$^{1.22}_{0.79}$ & \textbf{0.78}\,$^{0.95}_{0.62}$ & \textbf{0.66}\,$^{0.77}_{0.51}$ & 0.82\,$^{0.91}_{0.67}$ \\
 & GAFF2 & 1.11\,$^{1.39}_{0.81}$ & 0.81\,$^{1.02}_{0.62}$ & 0.62\,$^{0.73}_{0.50}$ & 0.81\,$^{0.89}_{0.68}$ \\
 & AceFF1.0 & 1.09\,$^{1.27}_{0.90}$ & 0.87\,$^{1.04}_{0.70}$ & 0.65\,$^{0.74}_{0.55}$ & \textbf{0.84}\,$^{0.90}_{0.73}$ \\
\midrule
JNK1 & AceFF-2-RESP-1 & 1.15\,$^{1.52}_{0.82}$ & 0.91\,$^{1.16}_{0.70}$ & \textbf{0.37}\,$^{0.59}_{0.11}$ & \textbf{0.49}\,$^{0.76}_{0.14}$ \\
 & GAFF2 & 1.13\,$^{1.42}_{0.84}$ & 0.89\,$^{1.13}_{0.67}$ & 0.34\,$^{0.59}_{0.05}$ & 0.45\,$^{0.74}_{0.08}$ \\
 & AceFF1.0 & \textbf{1.09}\,$^{1.31}_{0.84}$ & \textbf{0.86}\,$^{1.09}_{0.64}$ & \textbf{0.37}\,$^{0.58}_{0.13}$ & \textbf{0.49}\,$^{0.73}_{0.17}$ \\
\bottomrule
\end{tabular}
\end{table}

\begin{table}
\caption{As Table~\ref{tab:comp_ddg}, for absolute binding free energies $\Delta G$ recovered per ligand by MLE.}
\label{tab:comp_dg}
\begin{tabular}{llrrrr}
\toprule
& & \multicolumn{4}{c}{$\Delta G$ (kcal/mol)} \\
\cmidrule(lr){3-6}
System & Model & RMSE & MAE & $\tau$ & $\rho$ \\
\midrule
TYK2 & AceFF-2-RESP-1 & \textbf{0.34}\,$^{0.45}_{0.23}$ & \textbf{0.27}\,$^{0.38}_{0.17}$ & \textbf{0.85}\,$^{0.98}_{0.65}$ & \textbf{0.95}\,$^{1.00}_{0.80}$ \\
 & GAFF2 & 0.65\,$^{0.83}_{0.44}$ & 0.52\,$^{0.72}_{0.35}$ & 0.67\,$^{0.84}_{0.43}$ & 0.87\,$^{0.95}_{0.62}$ \\
 & AceFF1.0 & 0.74\,$^{0.92}_{0.53}$ & 0.59\,$^{0.81}_{0.38}$ & 0.80\,$^{0.96}_{0.58}$ & 0.91\,$^{0.99}_{0.69}$ \\
\midrule
CDK2 & AceFF-2-RESP-1 & 1.05\,$^{1.37}_{0.68}$ & 0.82\,$^{1.16}_{0.52}$ & 0.52\,$^{0.77}_{0.21}$ & 0.70\,$^{0.91}_{0.30}$ \\
 & GAFF2 & 1.03\,$^{1.37}_{0.64}$ & 0.79\,$^{1.14}_{0.50}$ & 0.45\,$^{0.82}_{0.11}$ & 0.51\,$^{0.91}_{0.02}$ \\
 & AceFF1.0 & \textbf{0.94}\,$^{1.32}_{0.45}$ & \textbf{0.64}\,$^{1.00}_{0.34}$ & \textbf{0.68}\,$^{0.87}_{0.40}$ & \textbf{0.87}\,$^{0.96}_{0.58}$ \\
\midrule
Thrombin & AceFF-2-RESP-1 & 0.83\,$^{1.12}_{0.51}$ & 0.66\,$^{0.97}_{0.38}$ & \textbf{0.56}\,$^{0.91}_{0.08}$ & \textbf{0.72}\,$^{0.97}_{0.11}$ \\
 & GAFF2 & 0.82\,$^{1.09}_{0.48}$ & \textbf{0.62}\,$^{0.95}_{0.32}$ & 0.55\,$^{0.98}_{-0.04}$ & 0.68\,$^{0.99}_{-0.00}$ \\
 & AceFF1.0 & \textbf{0.80}\,$^{1.10}_{0.47}$ & 0.64\,$^{0.95}_{0.38}$ & 0.42\,$^{0.84}_{-0.14}$ & 0.61\,$^{0.93}_{-0.10}$ \\
\midrule
p38 & AceFF-2-RESP-1 & \textbf{0.84}\,$^{1.06}_{0.61}$ & \textbf{0.65}\,$^{0.84}_{0.48}$ & 0.58\,$^{0.75}_{0.40}$ & 0.75\,$^{0.89}_{0.54}$ \\
 & GAFF2 & 0.99\,$^{1.30}_{0.68}$ & 0.75\,$^{0.98}_{0.55}$ & 0.64\,$^{0.79}_{0.45}$ & 0.79\,$^{0.92}_{0.58}$ \\
 & AceFF1.0 & 0.89\,$^{1.13}_{0.62}$ & 0.67\,$^{0.88}_{0.49}$ & \textbf{0.72}\,$^{0.82}_{0.59}$ & \textbf{0.89}\,$^{0.94}_{0.77}$ \\
\midrule
JNK1 & AceFF-2-RESP-1 & 1.13\,$^{1.40}_{0.86}$ & 0.95\,$^{1.22}_{0.69}$ & 0.49\,$^{0.71}_{0.22}$ & 0.71\,$^{0.88}_{0.37}$ \\
 & GAFF2 & \textbf{1.03}\,$^{1.26}_{0.76}$ & \textbf{0.83}\,$^{1.09}_{0.59}$ & 0.54\,$^{0.73}_{0.30}$ & 0.75\,$^{0.89}_{0.46}$ \\
 & AceFF1.0 & 1.15\,$^{1.48}_{0.78}$ & 0.92\,$^{1.22}_{0.64}$ & \textbf{0.55}\,$^{0.74}_{0.35}$ & \textbf{0.77}\,$^{0.88}_{0.54}$ \\
\bottomrule
\end{tabular}
\end{table}
Performance varies by target. For TYK2, electrostatic embedding with \texttt{AceFF-2-RESP-1} gives a consistent improvement across every accuracy metric compared to both classical GAFF2 and mechanical-embedding AceFF-1.0. TYK2 is a rigid benchmark system routinely used for method development~\cite{wang2015accurate, sabanes_zariquiey_quantumbind-rbfe_2025}. The size and consistency of this effect across all eight metrics suggests a real improvement rather than sampling noise. Errors fall by a factor of 1.7--2.2 against both baselines on every metric ($\Delta\Delta G$ RMSE $0.86 \rightarrow 0.45$ and $0.77 \rightarrow 0.45$~kcal/mol against GAFF2 and AceFF-1.0, respectively; MAE and both $\Delta G$ metrics follow the same pattern), and \texttt{AceFF-2-RESP-1} ranks first of the three methods on all eight metrics we report for this target. TYK2's binding pocket and ligand conformations are well-behaved and already well described by classical force fields, so static ligand charges were a primary limiting factor; replacing them with dynamic, conformation-dependent RESP charges closes this gap.

For the remaining four targets (CDK2, Thrombin, p38, and JNK1), electrostatic embedding gives neutral results, performing comparably to classical GAFF2 and mechanical-embedding AceFF-1.0. These systems are structurally more complex, with larger and more flexible ligands, flexible binding pockets, or net-charged species, and static ligand charges are likely not the main bottleneck here. One possible explanation is that coupling the MLIP charge model to the MM environment creates a more rugged potential energy landscape, raising the sampling cost and the statistical variance of edge $\Delta\Delta G$ values, consistent with the larger error bars on several of these edges. We cannot yet distinguish this from the simpler explanation that static charges were never the dominant error source for these four, more flexible targets, since target rigidity, size, and net charge all vary together across our five systems.
This is consistent with reports that MLIP-driven relative free energy calculations
show reduced repeatability for some potentials compared to classical MM~\cite{tkaczyk2026_repeatability_ani2x_maceoff},
and with the variance issues Rufa et al.~\cite{Rufa2020.07.29.227959} noted for
ML/MM reweighting.

\subsection{Correlation between $\Delta\Delta G$ errors and ligand MLIP force MAE}

\begin{figure}[htbp]
    \centering
    \includegraphics[width=0.8\textwidth]{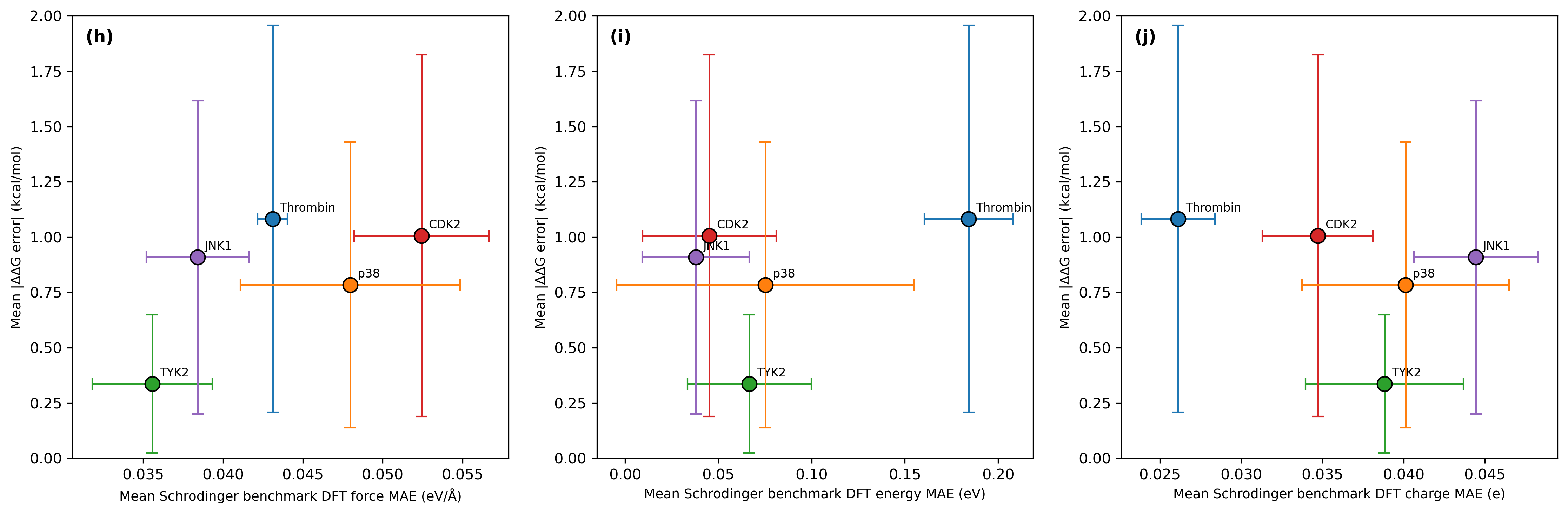}
    \caption{Correlation between per-target Schrödinger benchmark error metrics (force, energy, and RESP charge MAE, reported as $\text{mean} \pm \text{SD}$) and downstream mean absolute error in alchemical $\Delta\Delta G$ predictions across the five benchmark systems.}
    \label{fig:meta}
\end{figure}

To check whether standard single-molecule energy/force benchmark performance predicts downstream alchemical free energy accuracy, we compared Schrödinger benchmark errors against the observed RBFE outcomes across the five targets. We plotted the per-target mean and standard deviation ($\text{mean} \pm \text{SD}$) of the Schrödinger force, energy, and partial charge errors against the target-averaged $\Delta\Delta G$ absolute error, shown in Figure~\ref{fig:meta}.

Standard single-molecule energy/force benchmarks do not generally correlate with target-dependent RBFE outcome: energy and charge MAE on the Schrödinger dataset do not track which targets show electrostatic-embedding gains. TYK2, the only target with a clear improvement under electrostatic embedding, also has the lowest overall force MAE on the Schrödinger dataset; this pattern did not hold for the other four targets.

\subsection{Variation of partial charges}
\begin{figure}
    \centering
    \includegraphics[width=0.9\linewidth]{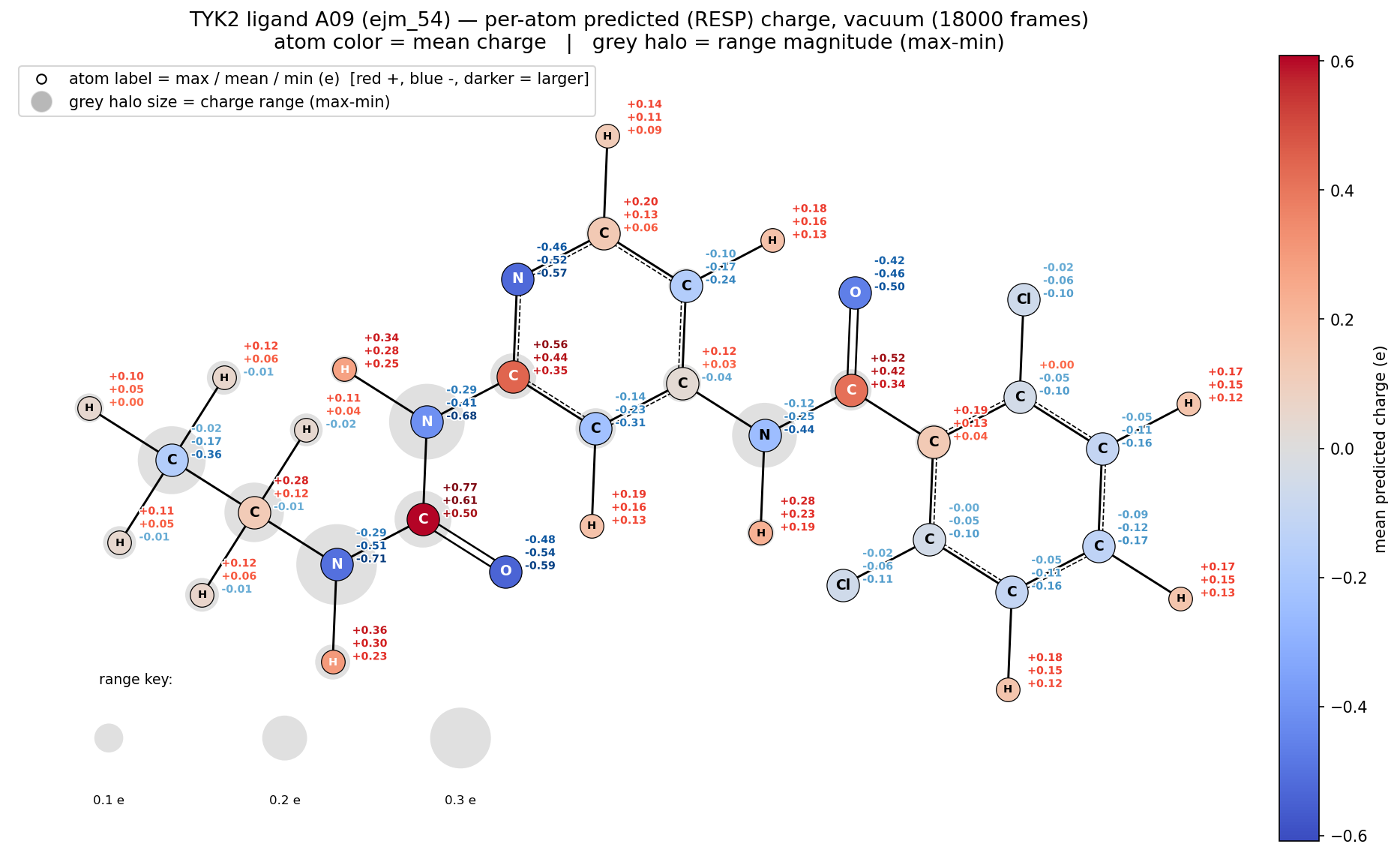}
    \caption{Variation of predicted partial charges for TYK2 ligand ejm54. A 2D depiction of the molecule is shown, the atoms are colored by the mean charge over the course of the simulation and the range of partial charge variation is depicted by the size of the shaded region and the labels.}
    \label{fig:lig_viz}
\end{figure}

To quantify how much the predicted partial charges vary over a simulation, we recorded the charge timeseries for one ligand in the TYK2 system. Figure~\ref{fig:lig_viz} shows a 2D depiction of the molecule, with each atom colored by its mean partial charge and labeled with the max, mean, and min charge; the shaded halo around each atom shows the range. Most atoms show small variation, under 0.1e. The largest range, 0.3e, is seen for some N atoms near the R groups. This magnitude of conformational charge variation is consistent with previous analyses of geometry-dependent charge models~\cite{semelak_advancing_2025, thurlemann2022_atomic_multipoles}.

\subsection{Limitations}

Several limitations bear on how far these results generalize. First, we fit RESP charges~\cite{bayly1993_resp} for compatibility with AMBER-family fixed-charge parameterizations, but Grassano et al.~\cite{grassano2024_embedding_schemes} found MBIS charges~\cite{verstraelen2016_mbis} to agree most closely with reference QM/MM embedding energies. We have not tested whether an MBIS-trained variant would perform better in the alchemical setting; the RESP choice is a compatibility argument, not an accuracy one.

Second, the embedding parameters were not optimized. Atomic polarizabilities come unchanged from the ANI-MBIS table~\cite{semelak_advancing_2025} where available, and from~\cite{vanduijnen1998_atomic_polarizabilities} otherwise; the effective dielectric constant is fixed at $\epsilon = 2$; and the Thole damping exponent~\cite{thole1981_damping} is set to $a = 1.3$. We have not checked how sensitive the computed free energies are to any of these three. Our short-range PME treatment also assumes that long-range electrostatics differ negligibly between RESP and AM1-BCC~\cite{jakalian2000_am1bcc_I} parameterizations.


Finally, we compare only against fixed-charge MM and mechanical embedding. Polarizable force fields such as AMOEBA and the Drude oscillator models~\cite{ponder2010_amoeba, lemkul2016_drude} target the same physics by a different route. 

\section{Conclusion}

We took the electrostatic embedding MLIP/MM formulation of Semelak et al.~\cite{semelak_advancing_2025} and retrained the charge model on RESP rather than MBIS charges, so it is commensurable with AMBER-family force fields. We built it on the TensorNet2 architecture~\cite{TensorNet, simeon_broadening_2025} as a single network predicting energies, forces, and charges (\texttt{AceFF-2-RESP-1}), and carried it through a full alchemical protein--ligand RBFE benchmark: five targets, three replicates per edge, matched against classical GAFF2 and mechanical-embedding AceFF-1.0 baselines on identical systems. This evaluation is the main result: it shows what the scheme does and does not buy in the setting it would actually be deployed in.

On TYK2, electrostatic embedding roughly halved the error against both GAFF2 and mechanical-embedding AceFF-1.0 ($\Delta\Delta G$ RMSE $0.86 \rightarrow 0.45\,\text{kcal/mol}$ and $0.77 \rightarrow 0.45\,\text{kcal/mol}$, respectively) and ranked first on all eight accuracy and correlation metrics, consistent with fixed baseline charges being the dominant error source on this target. On the four more flexible or complex pockets, electrostatic embedding performed comparably to the classical and mechanical-embedding baselines. We suggest this reflects dynamic charge variation creating a more rugged potential energy landscape that places greater demands on phase-space sampling, though with a single target showing improvement we cannot yet distinguish target rigidity from other differences between these systems.

Static benchmarks (energy or charge MAE on the Schrödinger dataset) do not directly correlate with target-dependent free energy accuracy. TYK2 combined good $\Delta\Delta G$ accuracy with the lowest force error on the Schrödinger benchmark; this pattern did not hold for the other four targets.

Future work will focus on exploring more sophisticated embedding schemes, particularly ones that do not leave the protein itself described by fixed, static charges. Anisotropic message-passing potentials such as AMP-BMS/MM~\cite{thurlemann2026ampbms} already extend electrostatic embedding to the protein side of the ligand--protein interaction, and adapting this kind of scheme to the alchemical RBFE setting is a natural next step.

\section*{Data and Software Availability}
\label{sec:data}
This work required modifications to two packages: TorchMD-Net and ATM. The other packages were used unchanged. The versions are listed in table~\ref{tab:software}.

\begin{table}[htbp]
\centering
\small
\begin{tabularx}{\textwidth}{@{}lllX@{}}
\toprule
Component & Role & Version / commit & Availability \\
\midrule
\multicolumn{4}{@{}l}{\textit{Modified in this work}} \\
AceFF-2-RESP-1 & Model weights & & \url{https://huggingface.co/Acellera/AceFF-2-RESP-1} \\
TorchMD-Net (fork)  & Model architecture     &  & \url{https://github.com/torchmd/torchmd-net/tree/resp_model} \\
ATM (fork)  & Electrostatic embedding MLIP/MM  & & \url{https://github.com/Acellera/atm/tree/electrostatic_embedding} \\
\midrule
\multicolumn{4}{@{}l}{\textit{Used unmodified}} \\
PySCF~\cite{sun_recent_2020}          & RESP labeling   & v2.10.0 & \url{https://pyscf.org} \\
GPU4PySCF~\cite{li2024introducting, wu2024enhancing} & GPU DFT & v1.4.2 & \url{https://github.com/pyscf/gpu4pyscf} \\
\texttt{cinnabar}~\cite{Hahnetal2022} & FE analysis & v1.0 & \url{https://github.com/OpenFreeEnergy/cinnabar} \\
AceFF-1.0~\cite{aceff_huggingface}    & mech.\ embedding baseline & commit \texttt{a8ad1ae}      & \url{https://huggingface.co/Acellera/AceFF-1.0} \\
\bottomrule
\end{tabularx}
\caption{Software used in this work, with the modified forks distinguished from
packages used as released.}
\label{tab:software}
\end{table}

The modifications are, in summary: in TorchMD-Net, the model that outputs the predicted RESP charges; and in Acellera's ATM package (itself based on the
Gallicchio-Lab AToM-OpenMM codebase~\cite{azimi2022_atm}) the code that implements the electrostatic embedding. \texttt{AceFF-2-RESP-1} weights are available from Acellera's Hugging Face page.

\section*{AI Tool Disclosure}
AI-based tools, including large language models, were used to assist with
editing the manuscript text and with implementing parts of the accompanying
analysis and simulation code. 

\begin{acknowledgement}
We acknowledge the Torres-Quevedo Program
(PTQ2023-012967/AEI/10.13039/501100011033).
\end{acknowledgement}

\section*{Competing Interests}
S.E.F. and G.D.F. have a potential conflict of interest due to direct interests in Acellera.

\bibliography{ref1, ref2, ref3}

\end{document}